\documentstyle[12pt]{article}
\begin{document}
\title{\Large \bf Dependence of Spiral Galaxy Distribution on Viewing Angle in RC3
\date{ }
\thanks{This work is supported by both the National Natural Science
Foundation of China under Grant 19603003, and K. C. Wong Education Foundation,
Hong Kong.}}

\author{{\small MA Jun$^{1,2,3,4,5}$,
SONG Guo-xuan$^{1,2,4}$ and SHU Cheng-gang$^{1,2,4}$}}
\maketitle
{\begin{center} $^1$ Shanghai Astronomical Observatory, Chinese
Acadenmy of Scienes,\\ Shanghai 200030

$^2$ National Astronomical Observatories, Chinese Acadenmy of Scienes,\\
Beijing 100012

$^3$ Chinese Academy of Science-Peking University Join Beijing
Astrophysical Center,\\
Beijing 100087

$^4$ Joint Lab of Optical Astronomy, Chinese Acadenmy of Scienes,\\
Shanghai 200030

$^5$ Young Astronomical Center, Shanghai Astronomical
Observatory,\\ Chinese Acadenmy of Scienes, Beijing 100871

E-mail: (majun, gxsong, cgshu)@center.shao.ac.cn
\end{center}}
\begin{abstract}
The normalized inclination
distributions are presented for the spiral galaxies in RC3.
The results show that,
except for the bin of $81^{\circ}$-$90^{\circ}$, in which
the apparent minor isophotal diameters that
are used to obtain the inclinations,
are affected by the central bulges,
the distributions for Sa, Sab, Scd and Sd are well consistent
with the Monte-Carlo simulation of random inclinations
within 3-$\sigma$, and Sb and Sbc almost,
but Sc is different.
One reason for the difference between the real distribution
and the Monte-Carlo simulation of Sc
may be that some quite inclined spirals,
the arms of which are inherently loosely wound on
the galactic plane and should
be classified to Sc galaxies, have been incorrectly
classified to the earlier ones,
because the tightness of spiral arms which is one of the criteria
of the Hubble classification in RC3 is
different between on the galactic plane and on the
tangent plane of the celestial sphere.
Our result also implies that there might exist biases
in the luminosity functions of individual Hubble types if
spiral galaxies are only classified visually.
\end{abstract}


A spiral galaxy inherently consists of a halo, bulge and thin disk with
spiral structure, which emerges from the central region
or the end of a bright bar. Optical images of spirals, which are
the projected ones on the tangent plane of the celestial sphere,
are dominated by the light
coming from stars, as modified by the extinction and reddening of dust.
If galaxies with the same morphologies
are oriented at different angles of inclination (i.e.,
the angle between the galactic
plane and the tangent plane of the celestial sphere), their images are
different. It means that the inclination of a spiral galaxy affects
its image, especially, the extent to which the arms are unwound. When
a spiral galaxy has a large inclination, its arms would appear
tightly coiled on the image even if they should be loosely
wound on the galactic plane.

Hubble$^{1, 2}$ introduced an early scheme to classify galaxies.
Its concepts are
still in use, which consists of a sequence starting from elliptical,
through lenticular,
to spiral galaxies. This scheme has been extended by some astronomers$^{3-14}$
over the years, who try to employ multiple classification criteria.
Galaxies originally classified Sc on the Hubble system cover a large
interval along the sequence, ranging from regular well-developed
arms in early Sc to nearly chaotic structures in very late
Sc. de Vaucouleurs$^{4, 5}$ introduced Scd, Sd, Sdm, Sm and Im
to Sc and SBc. In his classification, Sc class has well-developed
arms, and the arms become more and more chaotic
and the tightness of spiral arms is
not considered any more from Scd to Im.
Galaxy morphological classification is
still mainly done visually by
dedicated astronomers on the images, based on the Hubble's original scheme and
its modification. It is possible for each of observers to give slightly
different weights to various criteria, although the criteria for
classification are accepted generally by them. Lahav et al.$^{15}$
and Naim et al.$^{16}$ investigated the consistency of visual morphological
classifications of galaxies from six independent observers, and found
that individual observers agree with one another with combined rms
dispersions of between 1.3 and 2.3 type units, typically about 1.8 units
of the revised Hubble numerical type, although there are
some cases where the range of types given to it was more than 4 units in width.

In the present letter, we investigate how dependence of spiral galaxies
distribution on viewing angles is.
The sample adopted in this letter is from
the Third Reference Catalogue of Bright Galaxies by de Vaucoulours et al.$^{17}$
(hereafter, RC3),
the Hubble types of which are
from Sa to Sd. In RC3, it is complete
for galaxies that have apparent diameters at the $D_{25}$ isophotal level
larger than 1 arcminute
and the total B-band apparent magnitudes $B_T$
brighter than about 15.5, with a redshift not in excess of
15000 kms$^{-1}$.
In order to complete the analyzed sample
in the absolute magnitude, we select
the galaxies from RC3 that are larger than 1 arcminute
at the $D_{25}$ isophotal level, brighter than -20.1 for
the absolute B-band magnitudes and
limited  within a redshift of 10000 kms$^{-1}$.
It can be confirmed that the selected sample within
these ranges is complete in both luminosity and space.
The adopted values of
$d_{25}/D_{25}$ are from 0.2 to 1.0 (including 0.2 and 1.0) for
calculating the inclinations ($D_{25}$ and $d_{25}$
are the apparent major and minor isophotal
diameters measured at or reduced down to the surface brightness
level $\mu_B=25.0$ B magnitudes per square arcsecond in B-band.).
Our statistical sample contains 2519 spiral galaxies.
A Hubble constant of 75
kms$^{-1}$Mpc$^{-1}$ is assumed throughout.

The inclination of a spiral galaxy
is not only
an important parameter, but also difficult to determine.
If we assume that
the thickness of the spiral plane is rather inconsiderable
compared with its extension and, when a spiral galaxy
is inclined moderately to the plane of sky and the thickness
of nucleus can be omitted, the inclination may be
obtained by

\begin{equation}
\gamma=\arccos(\frac{d}{D}),
\end{equation}
where $D$ and $d$ are the apparent major and minor isophotal
diameters, respectively. When a spiral galaxy is seen quite edge-on, it is
not possible to count with the nuclear part as having a thickness
to be left out of consideration. Then, Eq.~(1) can not be
used to calculate the inclination, because
the apparent minor isophotal diameter consists of two parts,
one is attributed by the disk and, another by the bulge that
decreases the real value of inclination.
Considering that the disk is not infinitely thin,
Tully$^{18}$ corrected Eq.~(1) by

\begin{equation}
\gamma=\arccos\sqrt{[(\frac{d}{D})^{2}-0.2^2]/(1-0.2^2)}+3^{\circ}.
\end{equation}
The constant of $3^{\circ}$ is added in accordance with an empirical
recipe. But, when $d/D=0.2$, Eq.~(2) will give
a wrong inclination, that is $\gamma=93^{\circ}$.
The adopted formula for obtaining
the inclination of a spiral galaxy in the
present letter is

\begin{equation}
\gamma=\arccos\sqrt{[(\frac{d_{25}}{D_{25}})^{2}-0.2^2]/(1-0.2^2)}
\end{equation}
in order to avoid the inclination that is larger than $90^{\circ}$.

Figure 1 plots the normalized inclination
distribution for whole spiral galaxies
in our sample with bin size of $10^{\circ}$.
Open circles present individual values.
At the same time, we generate a Monte-Carlo sample of random
inclinations for comparison, which are presented by filled circles
with the error bars of
the solids and dots denoting 1-$\sigma$ and 3-$\sigma$ levels respectively.
From this picture, we can see that, except for the bin of
$81^{\circ}$-$90^{\circ}$, the distribution
is almost consistent with the Monte-Carlo random distribution within
3-$\sigma$.
This means that, in RC3, when the apparent minor isophotal
diameters are obtained, the effects of central bulges
have not been eliminated for some edge-on spiral galaxies.

Figs. 2-8 plot the normalized inclination distributions for Sa-Sd
galaxies with bin size of $10^{\circ}$, respectively. Open circles
also denote the values in each bin. In Figs. 2-8, the Monte-Carlo
samples of random inclinations are also produced and the error bars
have the same meanings as Fig. 1.
They show that,
except for the bin of $81^{\circ}$-$90^{\circ}$,
the distributions for Sa, Sab, Scd and Sd are well
consistent with the Monte-Carlo simulation of random inclinations
within 3-$\sigma$,
Sb and Sbc almost.
The distribution of Scd is consistent with
the Monte-Carlo simulation within 1-$\sigma$ very well.
But, the distribution of Sc is quite different from the
Monte-Carlo simulation.

Based on the analysis above, we find that, in RC3,
except for Scd and Sd, the central
bulges of which can be neglected,
the effects of central bulges
may have not been eliminated for some inclined spiral galaxies
when the apparent minor isophotal diameters are obtained.
At the same time,
the effects of central bulges only can not accout for
the difference between the real distribution
and the Monte-Carlo simulation for Sc,
and other factors are needed. A candidate may be that
some quite inclined spirals,
the arms of which are inherently loosely wound on
the galactic plane and should
be classified to Sc galaxies, have been incorrectly
classified to the earlier ones,
because the tightness of spiral arms, which is one of the criteria
of the Hubble classification in RC3, is
different between on the galactic plane and on the
tangent plane of the celestial sphere.
Our result also implies that there might exist biases
in the luminosity functions of individual Hubble types if
spiral galaxies are only classified visually.

{\noindent}{\bf Fig. 1.} Normalized inclination distribution for whole
sample spirals with bin size of $10^{\circ}$. The open and close circles
are individual values and a Monte-Carlo simulation of random
inclinations. The solid and dotted lines
denote 1-$\sigma$ and 3-$\sigma$ levels, respectively.

{\noindent}{\bf Fig. 2.} Same as Fig. 1, but for Sa galaxies.

{\noindent}{\bf Fig. 3.} Same as Fig. 1, but for Sab galaxies.

{\noindent}{\bf Fig. 4.} Same as Fig. 1, but for Sb galaxies.

{\noindent}{\bf Fig. 5.} Same as Fig. 1, but for Sbc galaxies.

{\noindent}{\bf Fig. 6.} Same as Fig. 1, but for Sc galaxies.

{\noindent}{\bf Fig. 7.} Same as Fig. 1, but for Scd galaxies.

{\noindent}{\bf Fig. 8.} Same as Fig. 1, but for Sd galaxies.

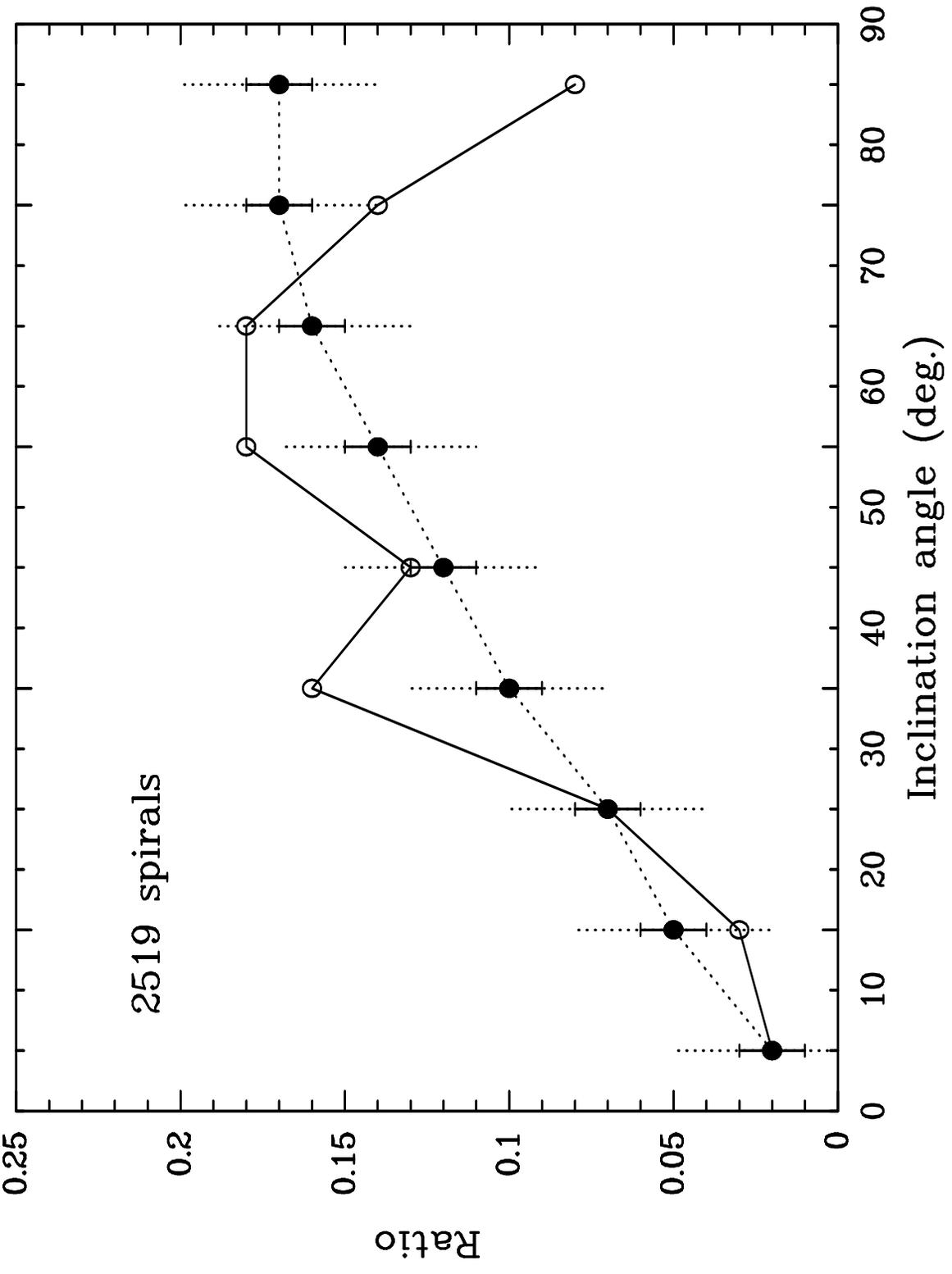
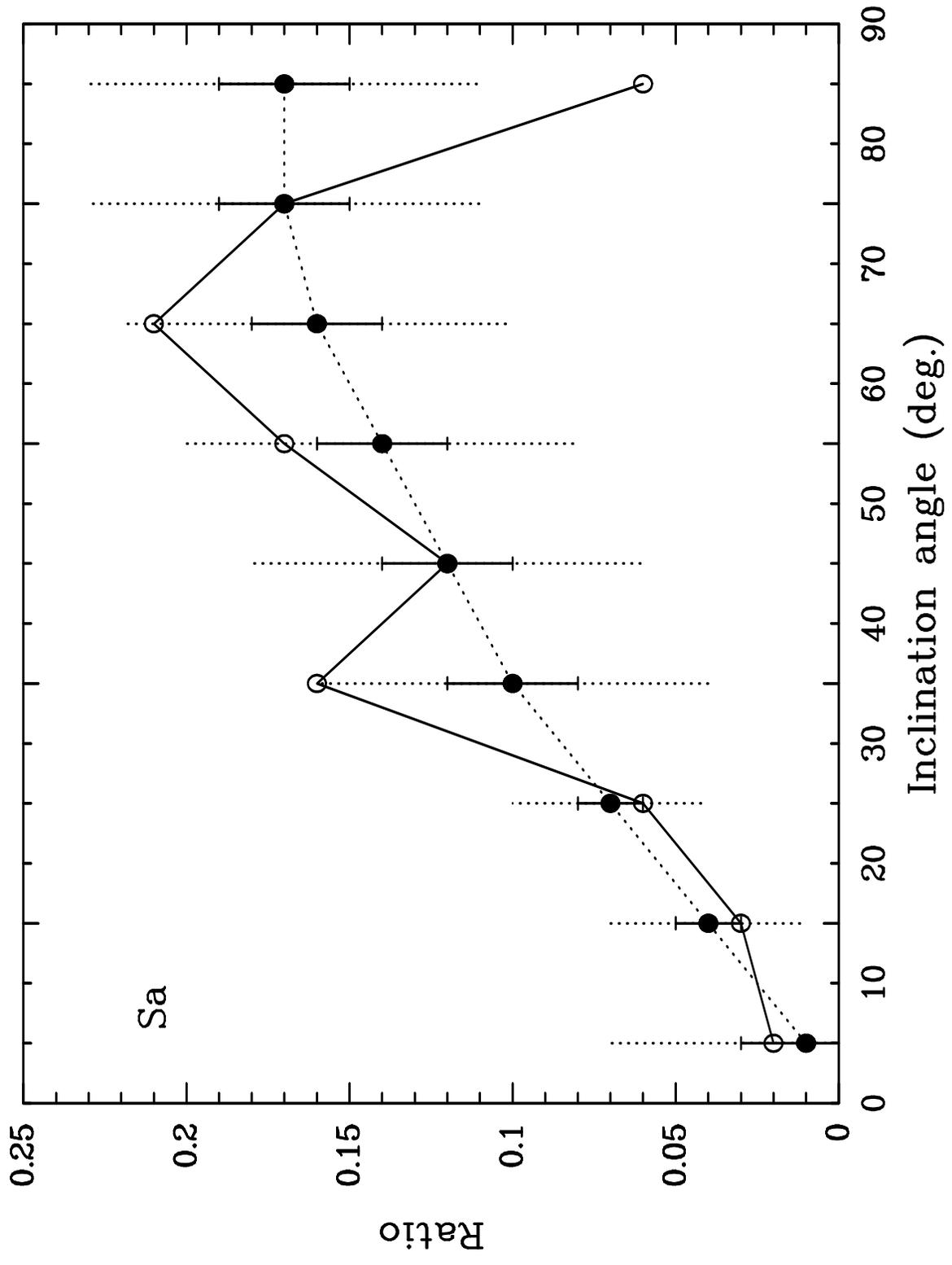
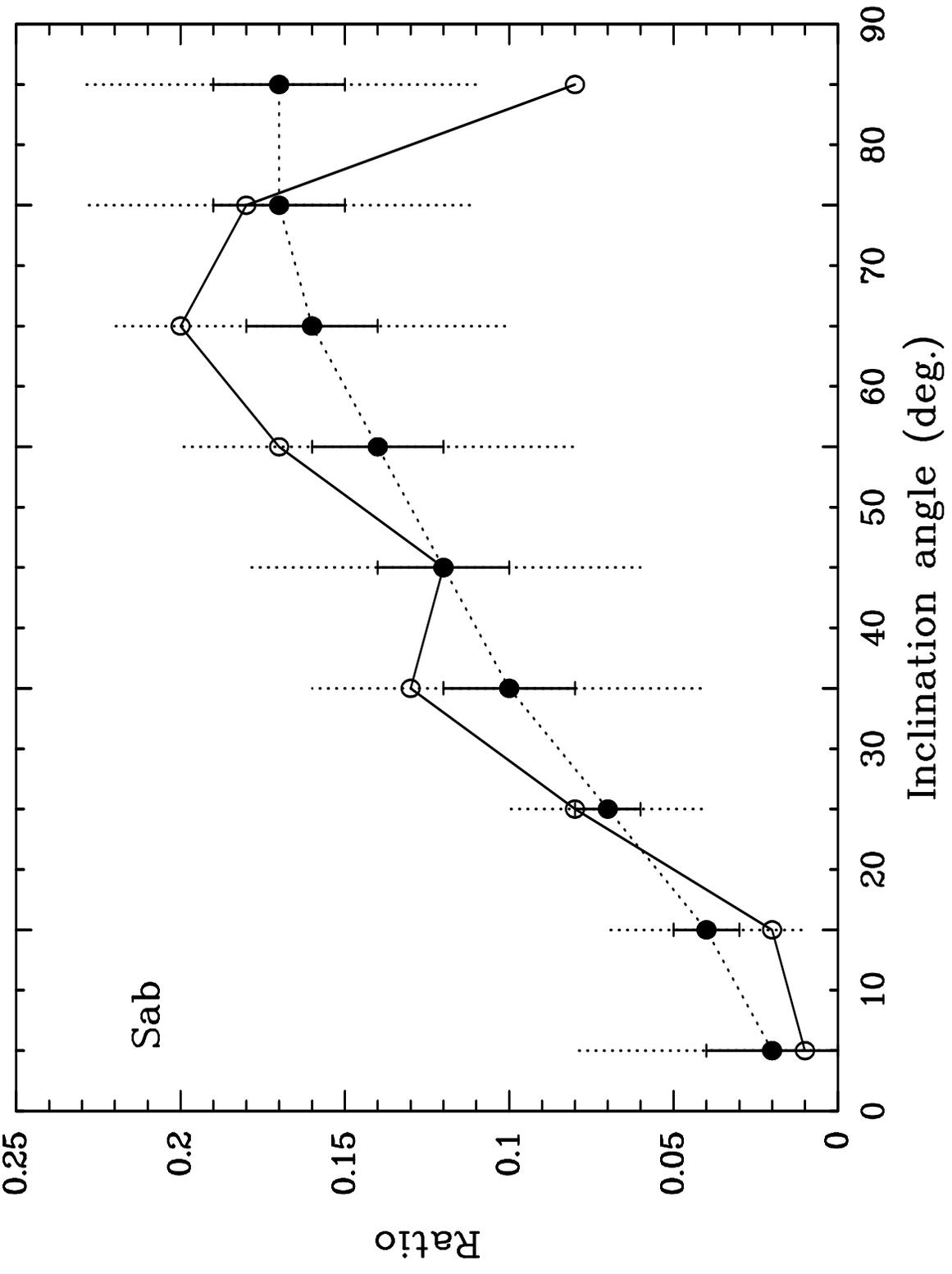
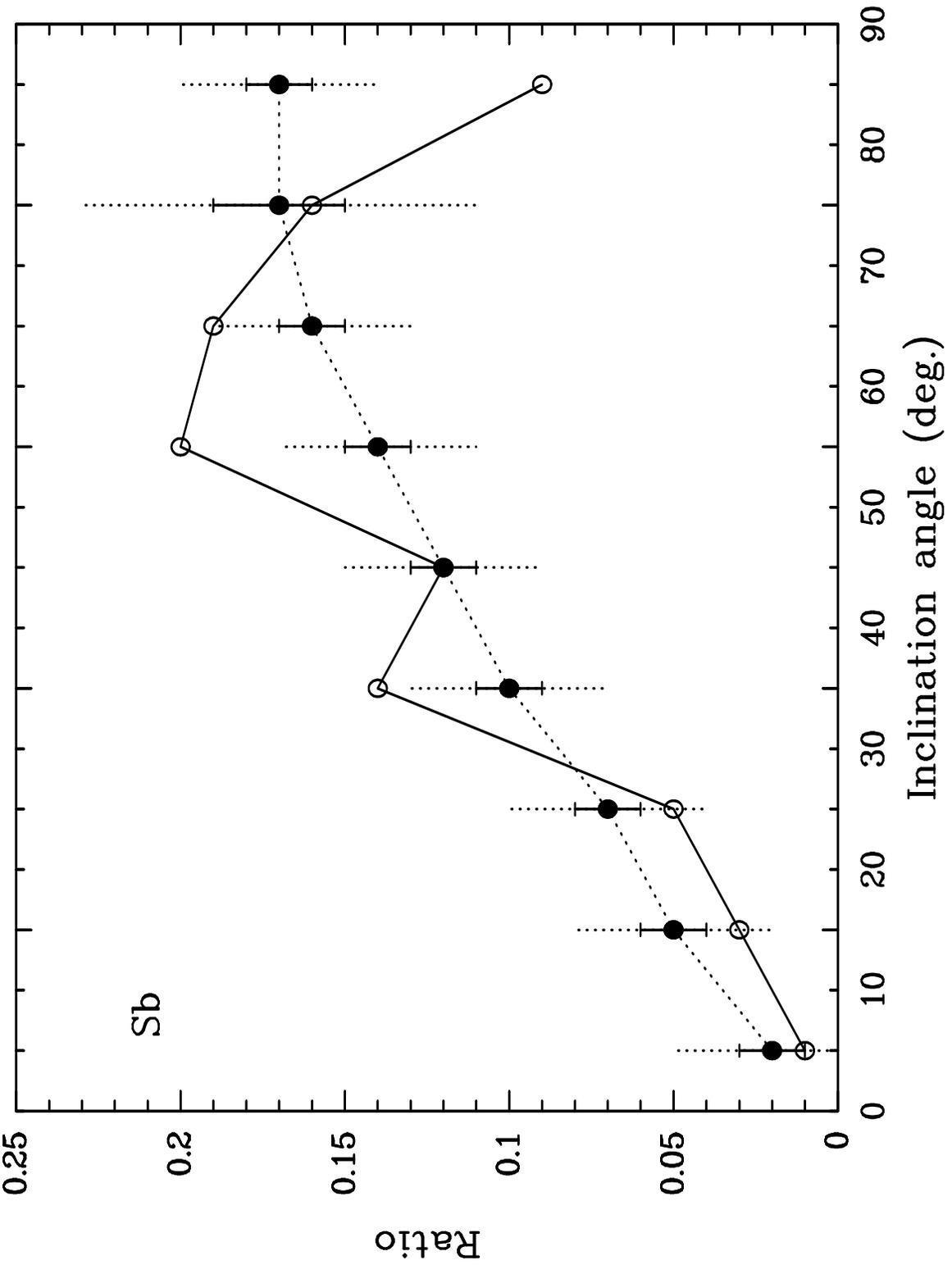
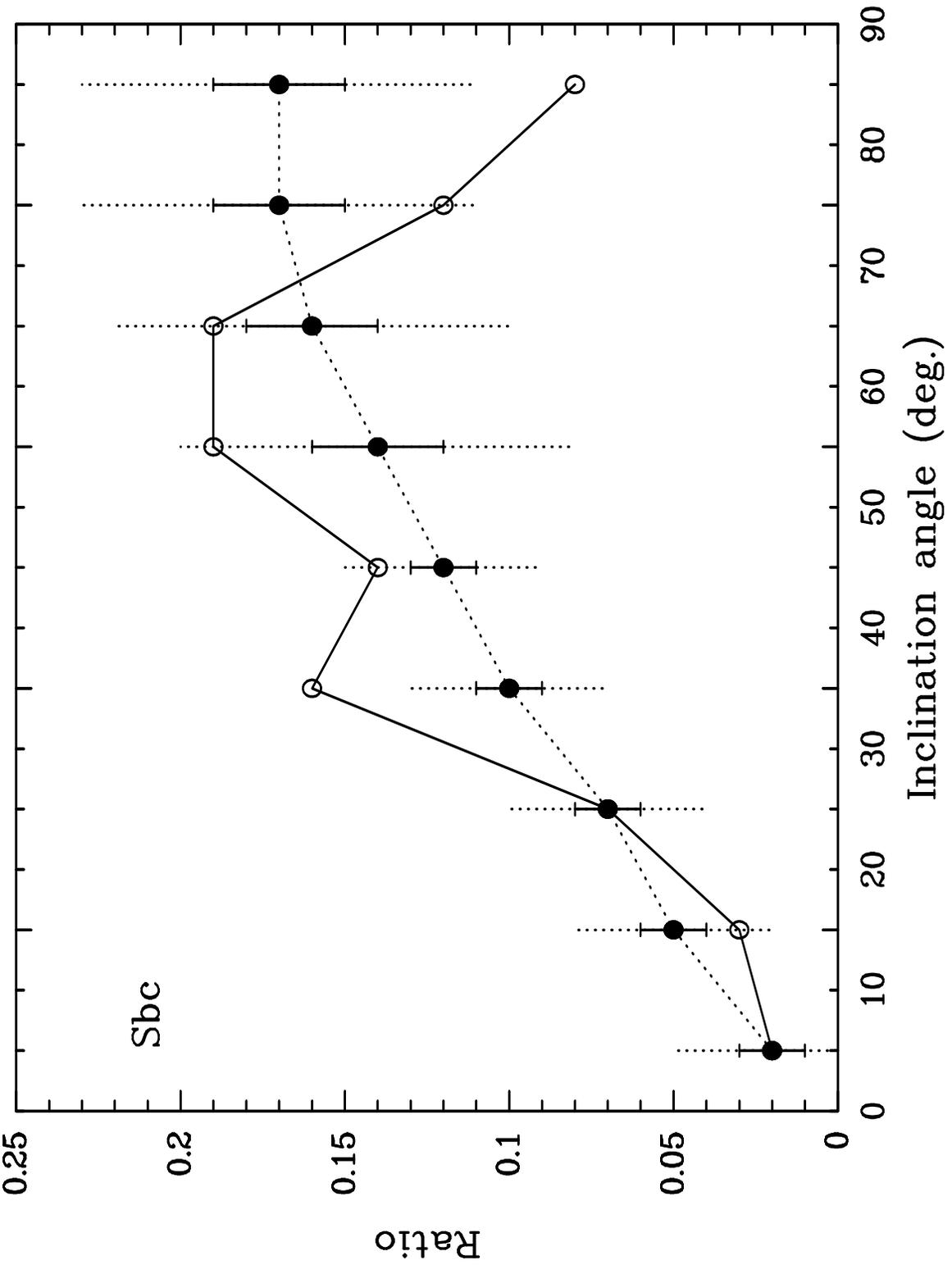
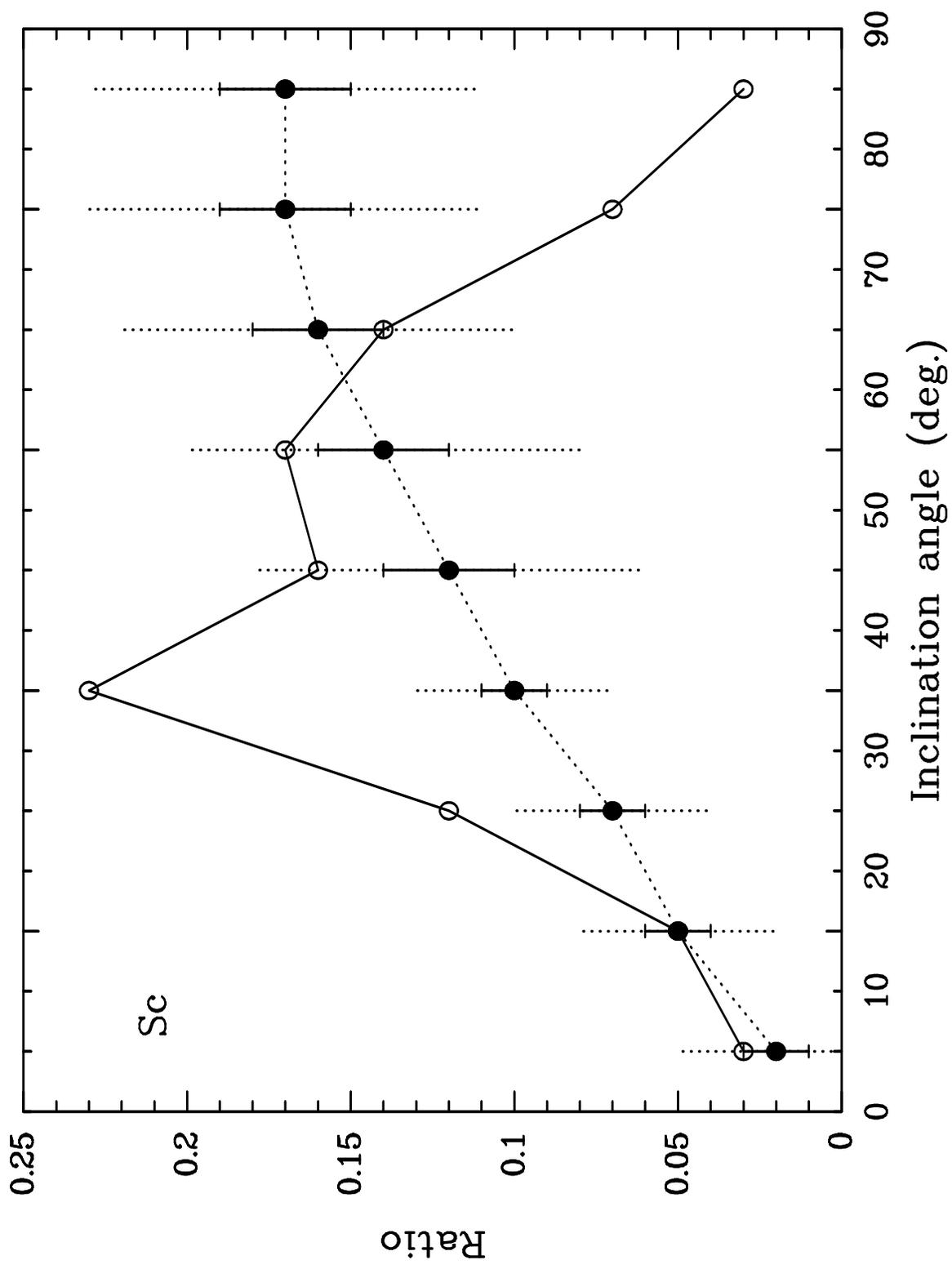
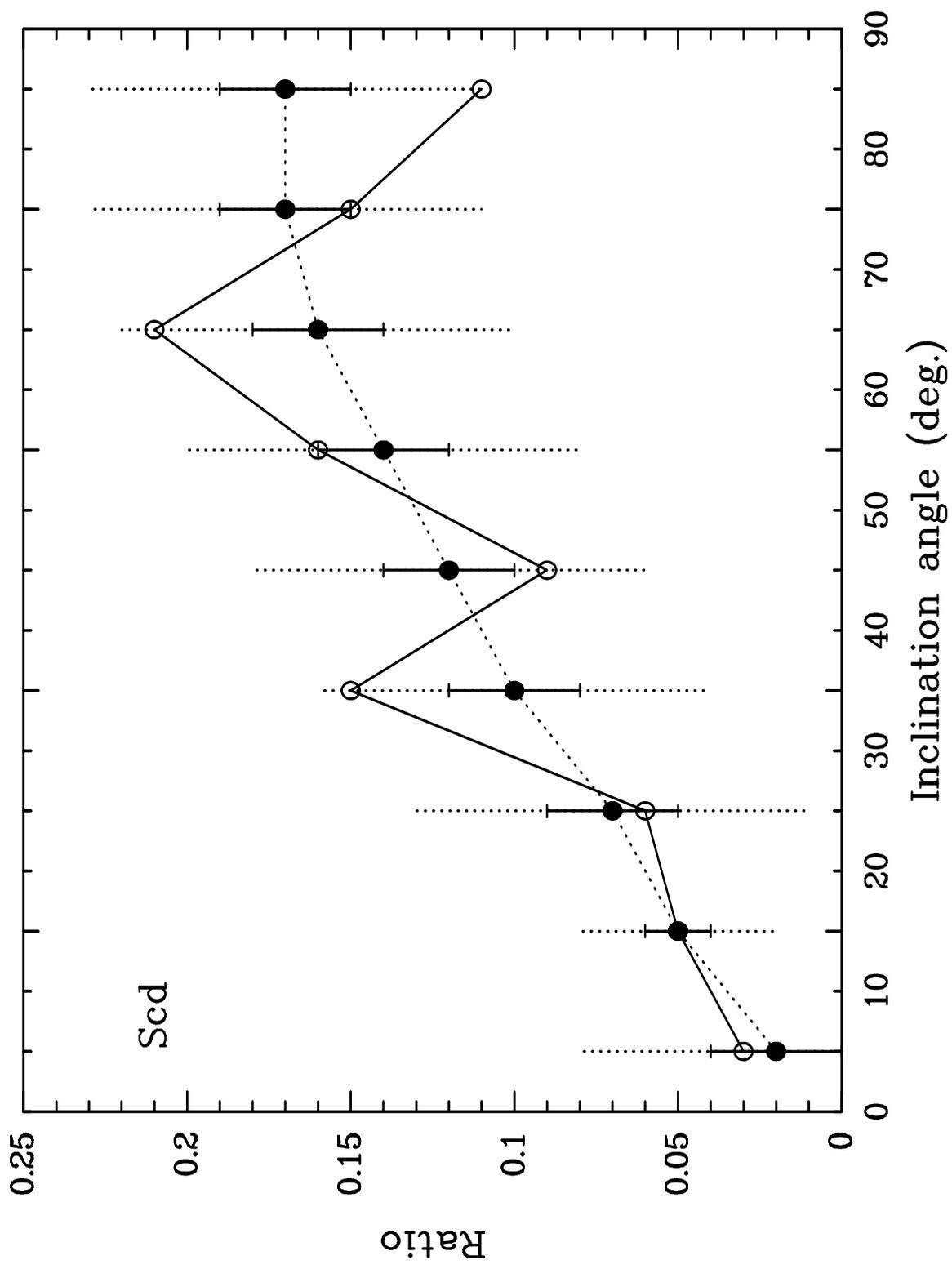
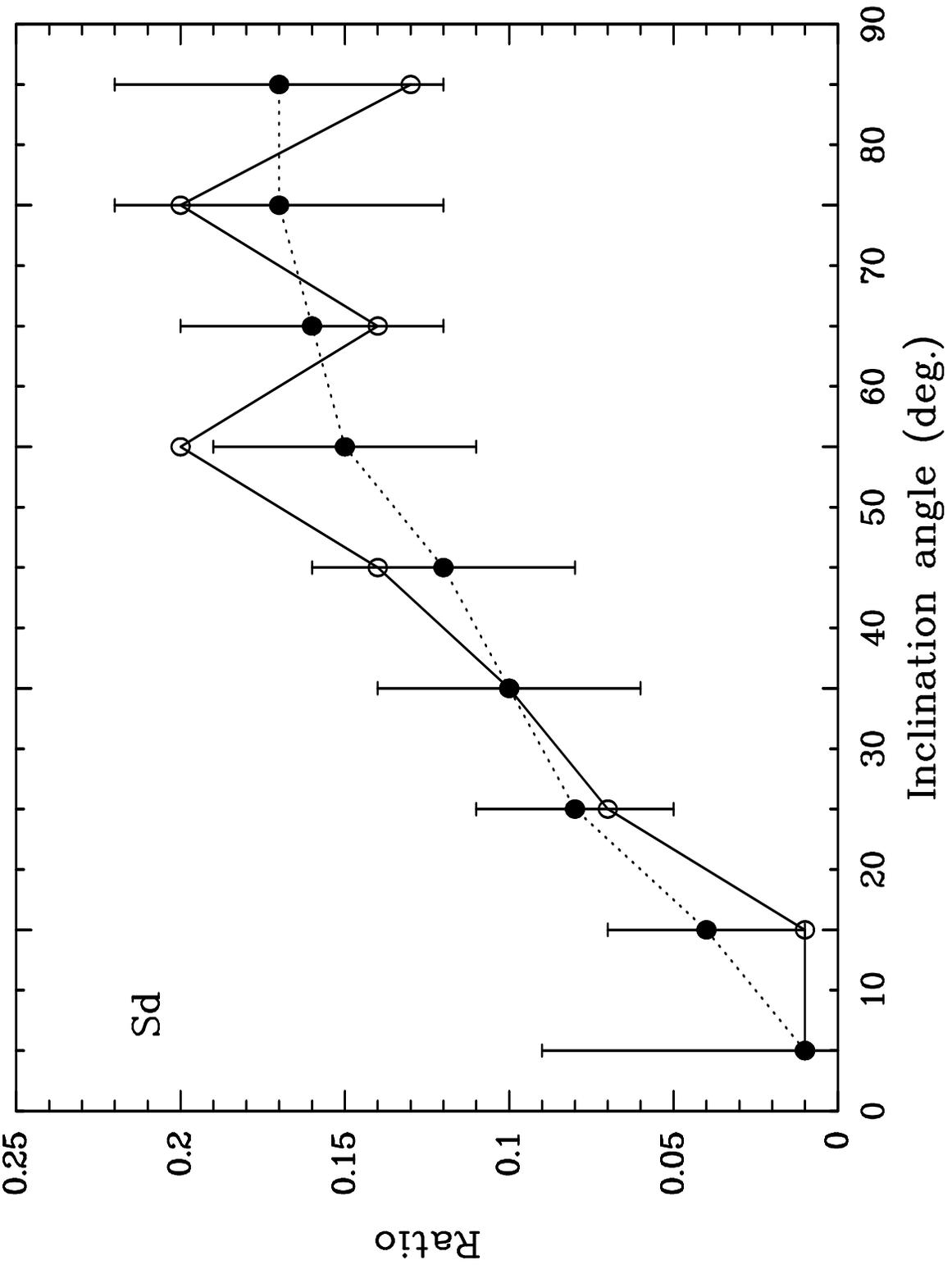
\end{document}